# Enhanced Raman scattering and weak localization in graphene deposited on GaN nanowires

Jakub Kierdaszuk,[1,*] Piotr Kaźmierczak,[1] Aneta Drabińska,[1] Krzysztof Korona,[1] Agnieszka Wołoś,[1,2]
Maria Kamińska,[1] Andrzej Wysmołek,[1] Iwona Pasternak,[3] Aleksandra Krajewska,[3,4] Krzysztof Pakuła,[1]
Zbigniew R. Zytkiewicz[2]

[1]Faculty of Physics, University of Warsaw, ul. Pasteura 5, 02-093, Warsaw, Poland
[2]Institute of Physics, Polish Academy of Sciences, Al. Lotnikow 32/46, 02-668, Warsaw, Poland
[3]Institute of Electronic Materials Technology, ul. Wólczyńska 133, 01-919, Warsaw, Poland
[4]Institute of Optoelectronics, Military University of Technology, ul. Gen. Sylwestra Kaliskiego 2, 01-476, Warsaw, Poland

The influence of GaN nanowires on the optical and electrical properties of graphene deposited on them was studied using Raman spectroscopy and a microwave-induced electron transport method. It was found that the interaction with the nanowires induces spectral changes and leads to a significant enhancement of the Raman scattering intensity. Surprisingly, the smallest enhancement (about 30-fold) was observed for the defect induced D' process and the highest intensity increase (over 50-fold) was found for the 2D transition. The observed energy shifts of the G and 2D bands allowed us to determine the carrier concentration fluctuations induced by the GaN nanowires. A comparison of the Raman scattering spatial intensity maps and the images obtained using a scanning electron microscope led to a conclusion that the vertically aligned GaN nanowires induce a homogenous strain, substantial spatial modulation of the carrier concentration in graphene and unexpected homogenous distribution of defects created by the interaction with the nanowires. The analysis of the D and D' peak intensity ratio showed that the interaction with the nanowires also changes the probability of scattering on different types of defects. The Raman studies were correlated with the weak localization effect measured using microwave-induced contactless electron transport. The temperature dependence of the weak localization signal showed electron-electron scattering as the main decoherence mechanism with an additional, temperature-independent scattering-reducing coherence length. We attributed it to the interaction of electrons in graphene with charges present on top of nanowires due to the spontaneous and piezoelectric polarization of GaN. Thus, nanowires act as antennas and generate an enhanced near field which can explain the observed significant enhancement of the Raman scattering intensity.

## I. INTRODUCTION

Graphene is considered an alternative material for use as an electrode in solar cells. Its main advantages are: high transparency (~98%),[1] high resistivity ($10^{-6}$ $\Omega$cm),[2] mechanical durability and flexibility.[3,4] Also recently it has been shown that a substantial increase in the solar cell efficiency can be achieved by using structures with nanowires which constitute straight pathways for respective electrodes to transfer separated carriers.[5] Gallium nitride in a wurtzite structure is a wide bandgap semiconductor with high spontaneous and piezoelectric polarizations along the c-axis.[6] In consequence, high polarization and free carriers of a high concentration could be present on the GaN surface.[7] Thus, when GaN nanowires covered with graphene on top are studied as a potential part of solar cell systems, it is essential to account for the influence of GaN nanowires on the graphene properties. Apart from the photovoltaic cells, the evaluation of the properties of the graphene/GaN NWs would be interesting from the point of view of graphene-based sensors. Graphene has been recognized as a promising candidate for sensing applications because of its superior electronic and mechanical properties.[2,3,4] In the case of gas sensors, molecules adsorbed on the graphene surface change the current in the transistor channel. The interaction with the external charges is also crucial for different graphene applications designed for working in aqueous solutions, like PH-meters or flow sensors.[8]

In this paper, graphene grown on a copper substrate and transferred onto gallium nitride nanowires was studied using Raman spectroscopy and microwave induced electron transport measurements. It was possible to trace how both surface topography and electric charges provided by the GaN NWs affect the strain and carrier distribution on the scale of hundreds of nanometers. The results were compared with those obtained for the reference structure, where graphene was deposited on a high-quality GaN epilayer. The reference sample independently showed interesting properties related to a different graphene behavior when lying either on the terraces or the edges of the GaN substrate.

Raman spectroscopy is one of the basic tools for graphene characterization. In addition to the G and 2D Raman peaks, related to the allowed inelastic light scattering processes, there are defect-activated D and D' peaks, corresponding to electron-phonon scattering processes in the vicinity of K and K' valleys (Dirac cones), and intra-valley processes, respectively. Therefore, the G and D peak intensity ratio provides information on the defect density,[9,10] while the D and D' intensity ratio depends on the defect type.[11]

One can expect two different mechanisms that lead to the modification of the observed scattering processes in graphene deposited on nanowires. Firstly, a direct contact with nanowires can create new defects and modify the properties of the defects already present in graphene. Secondly, the interaction with the substrate can substantially increase the probability of Raman scattering on selected defects and subsequently modify the observed ratio of scattering probabilities between different defect types in a way it happens in Surface Enhanced Raman Scattering (SERS) or Tip Enhanced Raman Scattering (TERS).[12] The SERS effect has been recently observed in graphene covered with gold nanoparticles, where the surface plasmons were induced by an incident electromagnetic field from metallic nanostructures.[13] The TERS mechanism was responsible for an enhancement in graphene located under the gold AFM tip.[14]

Another standard method employed to investigate scattering processes in different materials is magnetotransport. Among different variants of the magnetotransport methods, a novel contactless method based on microwave induced electron transport (MIET) is very attractive. This method can employ a standard electron spin resonance (ESR) spectrometer in quite a non-standard manner.[15] It probes microwave power reflected from the microwave cavity, which depends on the sample conductivity, and thus provides a non-invasive, contactless method of magnetoconductivity measurement. MIET allows one to probe a wide spectrum of processes in graphene, including the quantum interferences of electrons, which result either in weak antilocalization or weak localization depending on the scattering processes in graphene.[16,17] Weak antilocalization arises from negative quantum interferences which reduce the backscattering of carriers, resulting in the reduction of sample resistivity. These interferences occur because of the chiral nature of electrons in graphene with the Berry phase of $\pi$.[18] Additionally, since two non-equivalent Dirac cones are present in graphene, it is possible to change electron pseudo- and isospin, by for example scattering on defects. In this case, the positive quantum interferences leading to the increase of backscattering and sample resistivity can be observed.[17] This phenomenon is called weak localization. Graphene deposited on NWs is a very fragile system, which can be easily broken during the electric contact preparation. Therefore, a standard electron transport method used for weak localization or weak antilocalization measurements is very demanding technologically and as such can introduce some unwanted changes in the properties of the graphene/NWs structure. Thus, the MIET technique seems to be a very adequate method to study this system. It is worth stressing that Raman spectroscopy and MIET could link different electron scattering processes observed in graphene deposited on GaN nanowires and therefore provide new information about this system.

## II. EXPERIMENTAL DETAILS

In this paper, two types of samples were studied: the first one was graphene deposited on top of GaN NWs and the second one was a reference sample of graphene deposited on a high-quality GaN epilayer. The GaN NWs were grown by the Plasma Assisted Molecular Beam Epitaxy technique on a Si(111) substrate without the use of any catalyst and under N-rich conditions.[19] The GaN epilayer was grown using the metalorganic chemical vapour deposition (MOCVD) technique on a homoepitaxial substrate. The crystallographic orientation of both, the GaN epilayer and GaN nanowires was (000-1) (N-face).

Graphene was fabricated using the Chemical Vapour Deposition (CVD) technique on a copper substrate with methane gas as the precursor. The process of transferring graphene onto a GaN epilayer was performed by means of a standard method for substrates using poly(methyl methacrylate) (PMMA),[20] whereas for the GaN NWs a polymer-free transferring method was applied.[21] The width and length for both samples equalled several millimetres. The number of graphene layers was estimated by analysing the 2D peak width.[22,23] In the case of graphene deposited on the GaN nanowires less than 20% of the surface was covered with bilayer graphene, whereas in the case of graphene deposited on the GaN epilayer it was less than 10%. This however did not affect the discussion of results presented in this paper. Scanning electron microscope imaging was performed using an AURIGA CrossBeam Workstation (Carl Zeiss) microscope. The Raman spectroscopy analysis of graphene deposited on the NWs was carried out using a T64000 Horiba Jobin-Yvon spectrometer with an Nd:YAG laser operating at 532 nm wavelength as an excitation source. An objective with a magnification of 100 was used, which allowed us to obtain the spatial resolution of approximately 300 nm in diameter. Raman micro mapping was performed on a 4.5 μm x 4.5 μm area for graphene deposited on the NWs and a 1.4 μm x 3 μm area for graphene deposited on epitaxial GaN with the lowest step of piezoelectric motors equal to 100 nm. For the MIET measurements, a Bruker ELEXSYS E580 spectrometer was used. The spectrometer operates at a microwave frequency of 9.4 GHz (X-band) with a TE102 resonance cavity and it is equipped with a helium cryostat allowing one to reach temperatures down to 2 K. During the presented measurement, the microwave power was 0.15 mW and the modulation amplitude 0.1 mT.

## III. EXPERIMENTAL RESULTS

### A. Significant enhancement of Raman spectra

Raman spectroscopy results show that the intensity of the Raman spectra measured for graphene deposited on the GaN NWs is much higher (more than one order of magnitude) than for graphene deposited on the GaN epitaxial layer, which is shown in Figure 1. For statistical reasons, two-dimensional spatial mapping of the Raman spectra was performed on both samples: graphene deposited on the GaN NWs and graphene deposited on the GaN epitaxial layer. Interestingly, except for a substantial enhancement of the Raman spectra intensity for graphene deposited on the NWs, a non-homogeneous distribution of Raman scattering was observed for graphene on the GaN epilayer with an evident enhancement for graphene on the edges of the GaN microsteps as compared to the spectra measured for graphene on the GaN terraces.

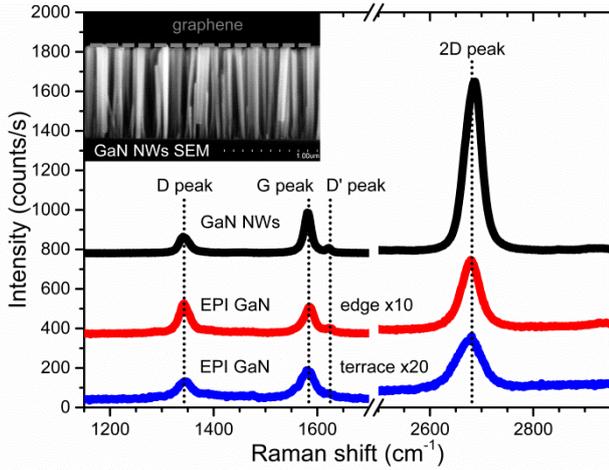

Figure 1. Raman spectra measured for graphene deposited on GaN NWs (black points), on microstep edge of GaN epitaxial layer (red points) and on terrace of GaN epitaxial layer (blue points). Note that spectra for graphene deposited on microstep edges and terraces of GaN epitaxial layer are multiplied by 10 and 20 respectively. Positions of the D, G, D' and 2D peaks were marked with black dashed vertical lines. In the inset: simple scheme of graphene layer deposited on NWs with GaN NWs SEM image.

Two-dimensional Raman mapping showed that the parameters of the Raman peaks (peak positions, widths and intensities) are changing locally with the topography of the NWs (changes in the D and G peak intensities are shown in Figures 2a and 2c). The observed spots are similar to the scanning electron microscope image of the surface of graphene deposited on the NWs presented in Figure 2b. Unfortunately, the light spot in Raman micromapping with a diameter of about 300 nm was larger than the average distance between the nanowires (about 100 nm). Therefore, the observed influence of the NWs on graphene is averaged over a few NWs. For graphene deposited on the epitaxial layer (see Figure 2d and 2f), the pattern corresponding to the GaN microsteps terraces and edges showed in Figure 2e is clearly observed.

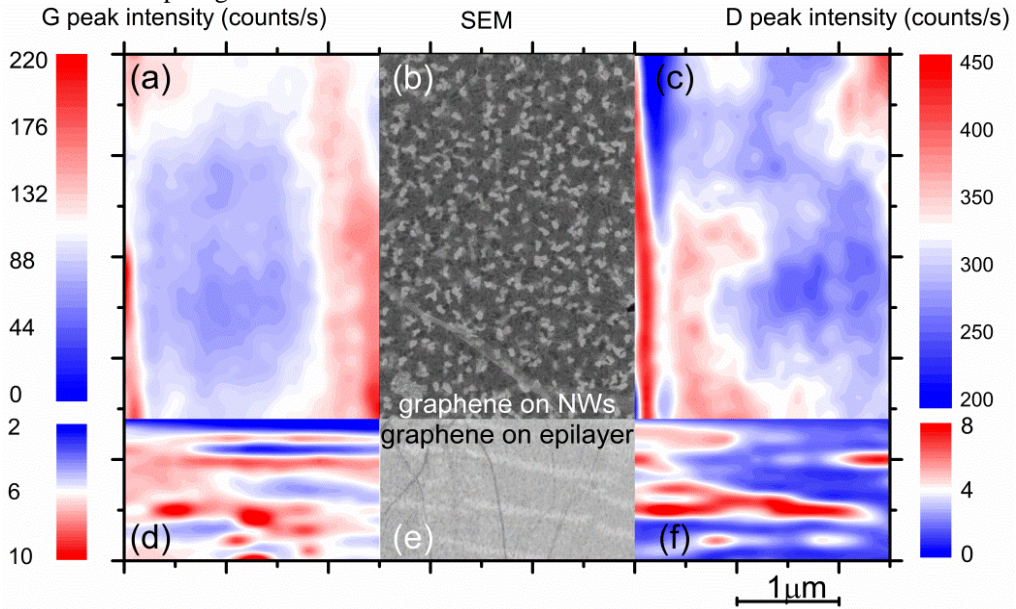

Figure 2. (a) two dimensional map of the G peak intensity for graphene deposited on NWs, (b) SEM image of the characteristic region of graphene deposited on NWs, (c) two dimensional map of the D peak intensity for graphene deposited on NWs, (d) two dimensional map of the G peak intensity for graphene deposited on epilayer, (e) SEM image of the characteristic region of graphene deposited on epilayer, and (f) two dimensional map of the D peak intensity for graphene deposited on epilayer.

Table 1. Average intensities of individual Raman peaks for graphene deposited on GaN NWs, graphene deposited on GaN epilayer microstep edges, graphene deposited on GaN epilayer terraces and enhancement ratios.

| Band | GaN NWs [1] | EPI GaN microstep edge [2] | EPI GaN terrace [3] | Ratio [1/3] | Ratio [2/3] | Ratio [1/2] |
|---|---|---|---|---|---|---|
| G  | 221.3 | 12.1 | 6    | 36.9 | 2   | 18.3 |
| 2D | 595.3 | 29.3 | 10.9 | 54.6 | 2.7 | 20.3 |
| D  | 110.2 | 12.8 | 2.4  | 45.9 | 5.3 | 8.6  |
| D' | 17.3  | 1.6  | 0.5  | 34.6 | 3.2 | 10.8 |

In Table 1, the average values of the D, G, 2D and D' peak intensities for graphene deposited on the NWs and graphene deposited on the epitaxial layer measured on the GaN microstep edges or terraces are presented.

The lowest intensities are recorded for peaks related to graphene on terraces and this will serve as a reference point in the vast majority of our discussion. However, some enhancement is observed for the spectra acquired on the GaN microstep edges and due to the possible accumulation of the polarization charge on the microstep edges a comparison with them will also sometimes be instructive. The intensity ratios of individual peaks show that the enhancement of the Raman spectra observed for graphene deposited on the NWs is more than 10 times stronger than the enhancement in the area of the GaN microstep edges for graphene deposited on the epilayer. However, it can be noticed that in the case of the microstep edges, the defect peaks (D and D') are mostly enhanced (3-5 times), whereas the G and 2D peak intensities increased 2-3 times. In the case of graphene deposited on the NWs, the 2D peak is the most enhanced peak with respect to graphene deposited on the GaN epilayer on terraces. It was observed that in the case of graphene deposited on the GaN NWs, the 2D and G peaks were enhanced about 20 times more, whereas the D and D' defect peaks were enhanced about 10 times when compared with the respective ones for graphene deposited on the GaN epilayer microsteps edges. Interestingly, the average intensity of the G peak on the GaN epilayer microsteps edges is slightly lower than that of the D peak, while for graphene on the GaN epilayer microsteps terraces and graphene on the NWs, the intensity of the D peak is about half of the intensity of the G peak. This indicates that in the case of graphene located on the GaN microsteps, scattering on defects is dominant, which is not the case for graphene deposited on the NWs and obviously for graphene on the GaN terraces.

It was reported that the Raman spectra for graphene covered with gold nanoparticles were enhanced by the SERS mechanism,[13] and were described with a theoretical model of enhancement in the system of nanoparticle and graphene hybrids, where the enhancement is proportional to the square Mie enhancement. For a nanowire, we can calculate the proportional enhancement ratio using the following equation:[13,24]

$$\frac{I}{I_0} \sim A = \left|\frac{\varepsilon(\omega)-1}{\varepsilon(\omega)+1}\right|^2, \qquad (1)$$

where $I$ is the enhanced band intensity, $I_0$ is the normal band intensity, $\varepsilon(\omega)$ is the dielectric function of the nanostructure material, $\omega$ is the frequency of the Raman band and $A$ is a proportional constant. Based on the dielectric function of gallium nitride,[25] we calculated $A$ for the D, G, D' and 2D bands. In contrast to the hybrids of gold nanoparticles and graphene, in the case of graphene on the GaN NWs, the highest enhancement was found for the 2D band, and the lowest enhancement was found for the D band.[26] Our experimental results show that the enhancement for the G and D' band intensities is smaller than for the D and 2D band intensities. Thus, despite the SERS mechanism, for graphene deposited on the NWs, the enhancement ratio is also changed by scattering on defects in graphene. Additionally, the height of individual NWs differs slightly (see the inset in Fig. 1), and therefore, also the TERS mechanism, with the enhancement depending on the distance from the individual nanowire tips to the graphene layer, has to be taken into consideration.[12]

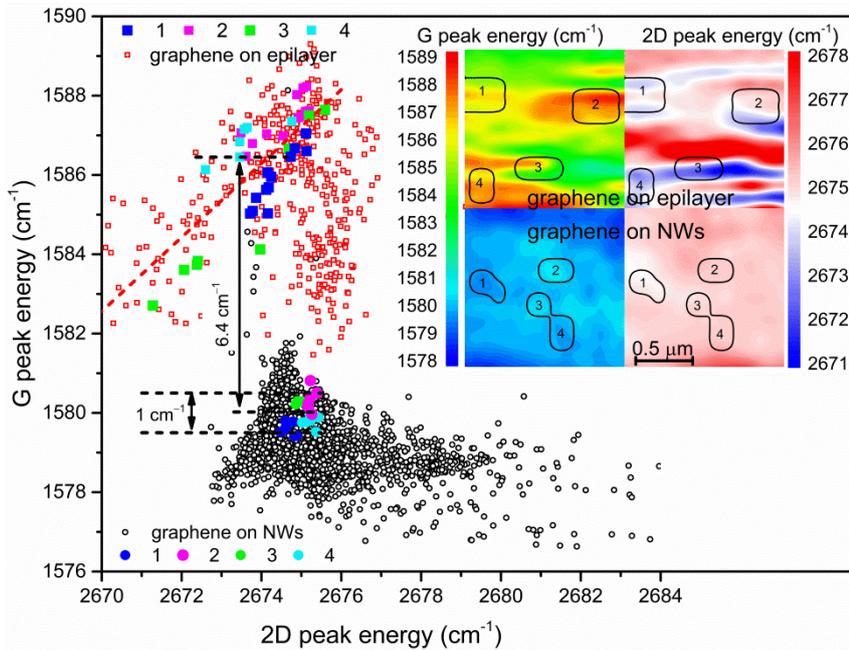

Figure 3. Mutual dependence of the G and 2D peak energies for graphene deposited on NWs (black points) and on GaN epilayer (red points) obtained from two dimensional Raman micromapping. Color points represent specific areas of the G and 2D peak energies in two dimensional Raman micromapping (inset). The red dashed line is the trend line with slope equal to 1.

To validate the SERS hypothesis a careful analysis of the energy positions of the Raman peaks was performed. The energy position of the G and 2D peaks allows us to extract information about the strain[27,28,29] and carrier concentration [27,30] of the graphene layers. The G peak energy ($E_G$) depends linearly on the carrier concentration, when the concentration of electrons or holes is lower than $2 \times 10^{13}$ cm$^{-2}$. A shift of the G and 2D peak energy can be also caused by the strain in the graphene structure.[27,31] We compared the dependence of the G peak energy as a function of the 2D peak energy $E_G(E_{2D})$ for both samples (Figure 3).

For graphene deposited on the epilayer, higher values of the G peak energy are observed as compared to graphene on the NWs. We can estimate the average changes of the carrier concentration using the following equation:[29]

$$\Delta n (\text{cm}^{-2}) = \frac{(E_{G1} - E_{G2})(\text{cm}^{-1})}{7.38\ (\text{cm})} \times 10^{13}. \quad (2)$$

The difference of the G peak energy between both samples is about 6.4 cm$^{-1}$, which corresponds to the carrier concentration difference of the order of $10^{13}$ cm$^{-2}$. To distinguish the correlation between the energies of the G and 2D peaks, we traced the $E_G(E_{2D})$ dependence for small areas for both samples (the inset in Figure 3). We observed substantial differences in the correlation between the energy of the G and 2D peaks for both samples. For graphene deposited on the epitaxial layer within the small region, large changes (up to 6 cm$^{-1}$) of both, G and 2D peak energies, were observed with a positive correlation and a slope coefficient of about 1. The simultaneous change of the strain and carrier concentration has to be considered to explain the observed correlation between the energies of the G and 2D peaks. Typical values reported in the literature for this coefficient for both the uniaxial and biaxial strain of graphene measured in experiments with external stresses applied to graphene were between 0.33 and 0.45.[27,32] These values are much lower than the ones observed for our sample. According to a simple model, where the 2D peak energy is independent of the concentration, the 2D band energy can be described as:[33]

$$E_{2D} = E_{2D}^0 - 2\gamma_{2D} E_{2D}^0 \varepsilon, \quad (3)$$

where $E_{2D}^0$ is the 2D peak energy of unstrained graphene (2677.6 cm$^{-1}$),[27] $\gamma_{2D}$ is the Grüneisen parameter characterizing the changes in 2D peaks phonon frequencies as a function of the crystal volume, $\varepsilon$ is the strain coefficient. A similar equation can be used to describe the G peak energy:[33]

$$E_G = E_G^0 - 2\gamma_G E_0^G \varepsilon + na, \quad (4)$$

where $E_G^0$ is the G peak energy of unstrained and undoped graphene (1583.5 cm$^{-1}$),[27] $\gamma_G$ is the G peak Grüneisen parameter, $n$ is the electron concentration and $a$ is a fit-derived slope parameter for the $E_G(n)$ dependence (7.38×10$^{13}$ cm).[29] Therefore, the slope coefficient of $E_G(E_{2D})$ can be calculated as $\frac{\gamma_G E_G^0}{\gamma_{2D} E_{2D}^0}$,[23] and furthermore, it gives the $\gamma_G/\gamma_{2D}$ ratio, which in the case of our sample is equal to 1.57. For a typical uniaxial and biaxial strain, the $\gamma_G/\gamma_{2D}$ ratio reported in the literature is lower than 1. This strongly suggests that the observed value of $\gamma_G/\gamma_{2D}$ is caused not only by the graphene strain but also by the change of the electron concentration in graphene induced by the electric field present in epitaxial gallium nitride. In the case of graphene deposited on the NWs, the 2D and G peak energies are nearly constant within the selected small regions. This shows that locally neither the strain nor the electron concentration are changed in graphene. On the other hand, regions where the 2D peak energy does not change while the G peak energy changes about 1 cm$^{-1}$ exist very close to one another, which corresponds to the electron concentration changes of the order of $10^{12}$ cm$^{-2}$ with a constant strain. These results show that the carrier concentration in graphene deposited on the NWs is locally modulated by the GaN nanowires. This redistribution can be induced either by polarization or free charges located on top of the GaN nanowires and the electric field induced by the spontaneous and piezoelectric polarization of gallium nitride. This result suggests that we are dealing with the Surface Enhancement Raman Scattering (SERS) mechanism, where the GaN nanowires act as an optical antenna and generate a strongly enhanced near field.[12] Since the GaN nanowires could change locally the Fermi level in graphene, the deposition of graphene on a set of randomly but uniformly distributed nanowires can induce a uniform modulation of the electric field and can be responsible for the observed significant enhancement of the Raman spectra in contrast to graphene deposited on the GaN epitaxial layer, where the electric field changes less rapidly and mostly on the microstep edges.

### B. Scattering on defects

Another important aspect that has to be considered is the problem of defects in graphene deposited on nanowires. As has been already mentioned, in addition to the allowed G and 2D Raman peaks, valuable information about both the concentration and types of defects in graphene can be obtained from the analysis of defect (D and D') peak intensities. It is commonly accepted that the ratio of the G peak intensity to the D peak intensity ($R_{GD}$) estimated by the maximum height of the Lorentz curve is inversely proportional to the defect density in graphene.[11] On the other hand, the ratio of the D peak intensity to the D' peak intensity ($R_{DD'}$) can provide information about defect types in graphene.[11,34] In the literature, different $R_{DD'}$ values have been assigned to different defect types. For example, the $R_{DD'}$ value of 1.3 was characteristic of on-site defects,[34] 3.5 indicated the occurrence of scattering at the boundaries in graphene,[11] 5 characterized a mix of single, double and complex vacancies,[35] 7 was characteristic of vacancies,[11] 10.5 was attributed to hopping defects[34] and 13 was related to sp$^3$ type defects.[11] On the other hand, calculations based on the density functional theory showed that the electric charges located outside the graphene surface give an undetectable contribution to the graphene Raman spectra.[34]

The obtained $R_{GD}$ ratio for graphene deposited on the nanowires (figure 4a) is on average lower than for graphene deposited on the epitaxial layer (figure 4c), which shows that graphene deposited on the nanowires is characterized by a higher defect density. Interestingly, the density of defects in graphene deposited on the NWs is homogenous and no clear nanowire-like pattern can be observed. In contrast, for graphene deposited on the epitaxial layer, a strip pattern characteristic of GaN terraces and macrostep edges can be observed with lower values of $R_{GD}$ near the microstep edges. By analysing the $R_{DD'}$ map for graphene deposited on the nanowires (fig. 4b, 5e), clear changes of the $R_{DD'}$ values correlated with the nanowire pattern are observed. There are spots with higher values of $R_{DD'}$ (about 11) surrounded by valleys with a small $R_{DD'}$ ratio (about 9) (fig. 5e). The fact that the Raman micromappings are correlated with the topography of the sample that result suggests that the most hopping defects are observed around individual nanowires. The interaction between the GaN NWs and graphene change the probability of scattering on different types of defects. Thus, the value of $R_{DD'}$ in graphene lying on a nanowire is different than the value in graphene hanging between nanowires. In the case of graphene deposited on the epitaxial layer, the D' peak is clearly visible only near the GaN microstep edges (fig. 4d), which limits the discussion of defect types to the area of step edges only. The histograms of the $R_{GD}$ and $R_{DD'}$ ratios and the $R_{DD'}(R_{GD})$ dependence for both samples are presented in fig. 5. A sum of lognormal distributions was fitted to $R_{GD}$ histograms and a sum of normal distributions was fitted to $R_{DD'}$ histograms. The $R_{GD}$ ratio for graphene deposited on the NWs has one distribution with a mean value of 0.4 and a standard deviation of 0.4 (denoted as d1 in fig. 5d). For graphene deposited on the epitaxial layer, two maxima are observed: the first with a mean value of 1.1 and a standard deviation of 0.2 and the second with a mean value of 3.1 and a standard deviation of 0.6 denoted as a1 and a2 in fig. 5a, respectively. This can be understood as the existence of two different defect types with an individual density in graphene deposited on the epilayer and one in graphene deposited on the NWs. It confirms the fact that in the case of graphene deposited on the epitaxial layer, the defect density is lower but the defect distribution is less homogenous than for graphene deposited on the NWs.

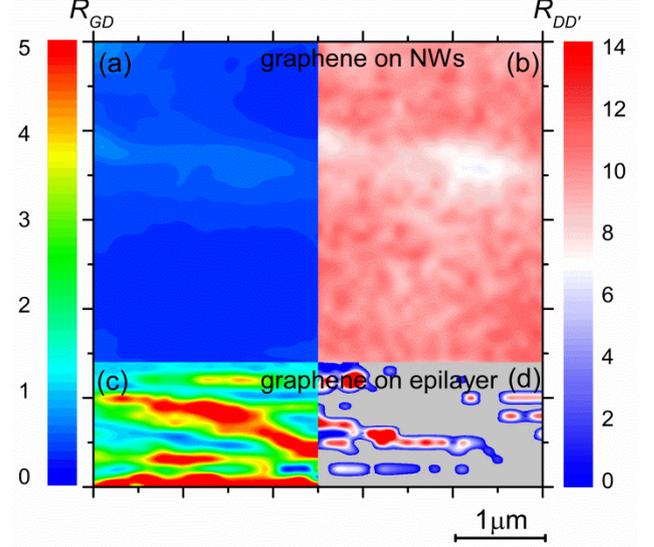

Figure 4. Two dimensional maps of (a) $R_{GD}$ ratio for graphene deposited on NWs, (b) $R_{DD'}$ ratio for graphene deposited on NWs, (c) $R_{GD}$ ratio for graphene deposited on epilayer, and (d) $R_{DD'}$ ratio for graphene deposited on epilayer.

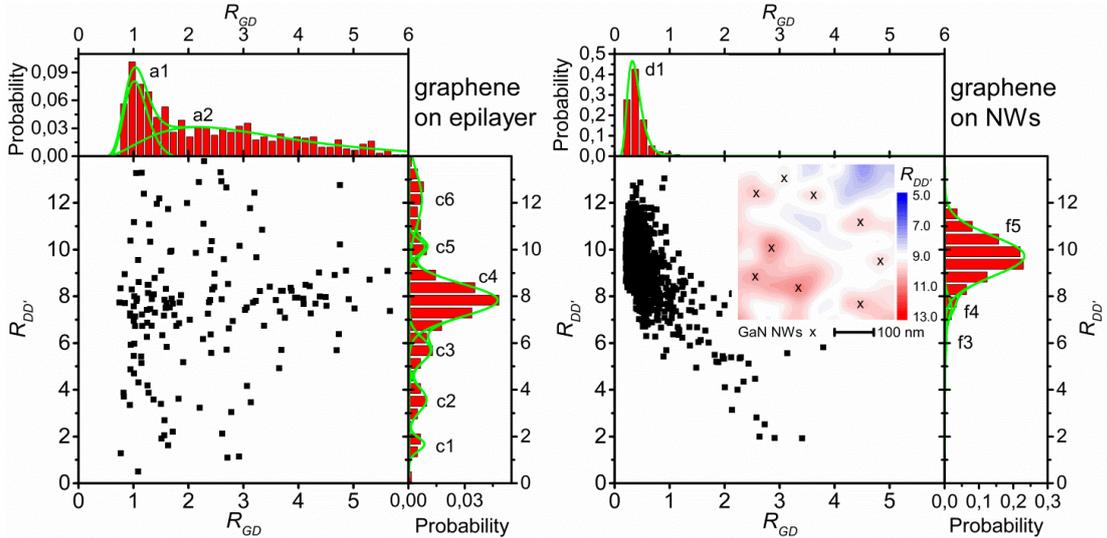

Figure 5. Histograms of RGD and RDD' ratios and correlation between them for both samples: (a) histogram of RGD for graphene deposited on epilayer with two maxima (a1, a2) and two fitted lognormal distributions, (b) correlation of RGD and RDD' ratios for graphene deposited on epilayer, (c) histogram of RDD' for graphene deposited on epilayer with six maxima (Tab. 2) and six fitted normal distributions, (d) histogram of RGD for graphene deposited on NWs with one maximum (d1) and one fitted lognormal distribution, (e) correlation of RGD and RDD' ratios for graphene deposited on NWs, (f) histogram of RDD' for graphene deposited on NWs with three maxima (Tab. 2) and three fitted normal distributions. Green dotted lines in the histograms are the fitted individual distributions. Green solid line is a sum of the fitted distributions. Two dimensional map of RDD' ratio of graphene on NWs with marked approximate position of GaN NWs is presented in the inset.

The analysis of the $R_{DD'}$ ratio shows differences in defect types between the samples. For graphene deposited on the epitaxial layer the D' peak could be observed only near the microstep edges, which corresponded to 26% of all collected spectra for graphene deposited on the epitaxial layer (fig. 4d). Six distribution maxima were found in the $R_{DD'}$ histogram (fig. 5c). These maxima were identified as the presence of on-site defects, scattering at the boundaries, mix of vacancies, vacancies, hopping defects and sp$^3$ defects (tab. 2).[11] The most abundant peak is related to $R_{DD'}$ equal to 7.8 and gives the total probability of vacancy type defects of 15.5% with respect to all collected spectra. The probability of other peaks is about 2% (tab. 2).

Table 2. Identification of defect types present in both samples, with probabilities of each observed defect type.

| $R_{DD'}$ value | Type of defect | Probability for EPI GaN | Probability for GaN NWs |
|---|---|---|---|
| 1.7 | on-site | 1.2% | - |
| 3.6 | boundaries | 2.4% | - |
| 6 | mixture of vacancies | 2.9% | 2.1% |
| 7.7 | vacancies | 15.5% | 4.2% |
| 9.9 | hopping defects | 1.1% | 93.5% |
| 12.1 | sp3 hybridisation | 3.4% | - |

Large redistribution of defect types for graphene deposited on the NWs was observed. The maxima related only to hopping defects, vacancies and a mixture of vacancies were detected. The probability and intensity of the maxima related to the vacancies and the mixture of vacancies are similar for both samples and those presented in table 2 (note that in the RDD' histogram for graphene on the NWs (fig. 5f), the scale is five times larger than in the RDD' histogram of graphene on the epilayer (fig. 5c)). Interestingly, the probability of hopping defects (defects that distort the bonds between the carbon atoms, retaining the number of carbon atoms) in graphene on the NWs is 85 times larger than for graphene on the epilayer (tab. 2) with the RDD' value corresponding to the maximum of hopping defects distribution located around individual nanowires (the inset in fig. 5e). This, together with the absence of three maxima for graphene deposited on the NWs, can be understood as both: defect redistribution and creation of new defects introduced by a mechanical contact with the GaN nanowires and the interaction with charges located on top of the GaN nanowires.

No explicit correlation between the $R_{GD}$ ratio and the $R_{DD'}$ ratio was found for graphene deposited on the epitaxial layer (fig. 5b). About 65% of the measured spectra were located in the second maximum of the $R_{GD}$ ratio (a2 in fig. 5a) and corresponded to vacancies like defects. Lower $R_{GD}$ values from the a1 maximum were rather uniformly distributed over all possible $R_{DD'}$ values.

For graphene deposited on the NWs, the dependence of the $R_{GD}$ ratio on the $R_{DD'}$ ratio shows a decreasing correlation (fig. 5e). Most of the points (about 99%) from the maximum related to hopping defects ($R_{DD'}$ of about 10) are located in the main maximum of the $R_{GD}$ ratio and only a small number of points are located in the tail of the lognormal distribution in the $R_{GD}$ ratio (fig. 5d). The D' peak was not observed only in 0.3% of the measured spectra. A two dimensional map of the $R_{DD'}$ ratio of graphene on the NWs is showed in the inset of figure 5e. Spots with high values of the $R_{DD'}$ (~11) ratio correspond to the NWs positions and are surrounded by valleys with a lower $R_{DD'}$ ratio (~7). Therefore, the NWs strongly modulate the $R_{DD'}$ ratio with a constant $R_{GD}$ ratio.

The information about defect scattering obtained from the Raman experiments should be reflected in the electron transport properties of the investigated material. As it has been already mentioned, the contactless MIET measurements based on the changes of the microwave-cavity $Q$ factor (quality factor) with the applied magnetic field ($dQ/dB$) in an electron spin resonance (ESR) spectrometer were employed instead of standard electron transport methods.[30] The quality factor of the resonator can vary for different reasons. In the case of a classical ESR measurement, changes in the cavity $Q$ factor due to energy absorption ($A$) by spins are recorded. The absorption of microwave radiation leads to the magnetization change and the signal is proportional to $dA/dB$ in a typical ESR spectrometer. The $Q$ factor of the cavity is also sensitive to the variation of the sample resistivity. In this case, the signal recorded by the ESR spectrometer is proportional to the derivative $d\sigma/dB$, where $\sigma$ is the sample conductivity.[36]

The results of the MIET measurements obtained for graphene deposited on the GaN NWs and the GaN epilayer are presented in fig. 6a and 6b, respectively. For graphene deposited on the GaN NWs, the conductivity changes at low magnetic fields are smaller and broader than those for graphene deposited on the GaN epilayer. The observed behaviour could be explained in terms of conductivity suppression induced by positive quantum interferences, which leads to the enhanced backscattering probability. This phenomenon is called weak localization. The derivative of magnetoconductivity on the magnetic field due to weak localization can be described using the following equation:[15, 17]

$$\frac{d\sigma(B)}{dB} = \frac{e^2}{\pi h}\left[\frac{1}{B_\varphi}F'\left(\frac{B}{B_\varphi}\right) - \frac{1}{B_\varphi + 2B_i}F'\left(\frac{B}{B_\varphi + 2B_i}\right) - \frac{2}{B_\varphi + B_i + B_{lr}}F'\left(\frac{B}{B_\varphi + B_i + B_{lr}}\right)\right], \quad (5)$$

where

$$F'(B) = \frac{1}{B} + \psi'\left(\frac{1}{2} + \frac{1}{B}\right)\frac{1}{B^2}, \quad (6)$$

and $\psi'$ is trigamma function. Using equation:

$$L_{\varphi,i,lr} = \sqrt{\frac{h}{8\pi e B_{\varphi,i,lr}}} \quad (7)$$

we calculated the coherence length ($L_\varphi$), the elastic intervalley scattering length ($L_i$), and the elastic long range (intravalley and warping) scattering length ($L_{lr}$) for both samples. The intervalley and intravalley scattering processes activate the D and D' graphene Raman defect bands by providing the missing momentum by defects for both samples.

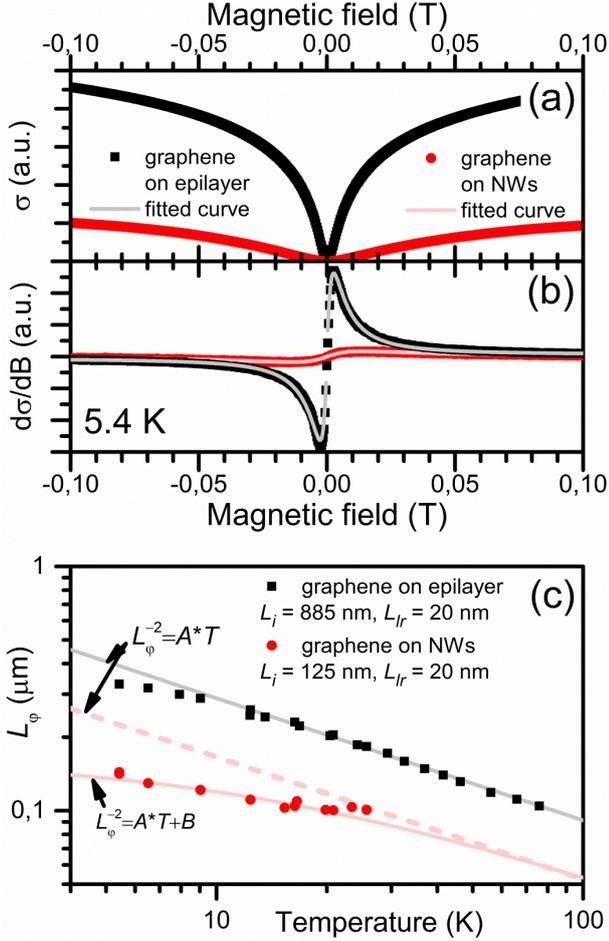

Figure 6. (a) Integrated measured signal proportional to conductivity for graphene deposited on NWs (black squares) and graphene deposited on epilayer (red circles), (b) measured derivative of conductivity on magnetic field for graphene deposited on NWs (black squares) and graphene deposited on epilayer (red circles) with fitted WL signal (grey and light red lines) using eq. (5), (c) temperature dependence of coherence length for graphene deposited on NWs (black squares) and epilayer (red circles) with fitted linear dependence of $L_\varphi^{-2}$ (T) with (light red solid line) and without ( grey and light red dashed lines) offset.

It was found that the elastic scattering lengths are independent of the temperature, whereas the coherence length decreased with temperature rapidly (fig. 6c). The long range scattering length values $L_{lr}$ were found to be the shortest. These were comparable for both samples (about 20 nm). On the other hand, the intervalley scattering length was found to be about 125 nm for graphene deposited on the NWs and was strongly reduced as compared to the value of 884 nm obtained for graphene deposited on the epitaxial layer. A similar trend was observed for coherence lengths. It was found that for graphene deposited on the NWs, the coherence length $L_\varphi$ decreased from about 150 nm at 5 K to 100 nm at 20 K, whereas for graphene deposited on the epitaxial layer the

$L_\varphi$ parameter decreased from 330 nm at 5 K to 104 nm at 80 K (fig. 5c). It is worth noting that the coherence length at 20 K obtained for graphene deposited on the epitaxial layer was about 200 nm, which is still a two-fold increase when compared with graphene deposited on the GaN NWs. The linear dependence of $L_\varphi^{-2}(T)$ suggested electron-electron scattering in the diffusive regime as the main decoherence mechanism for both investigated samples.[37] However, for graphene deposited on the NWs, a nonzero offset is clearly present, which could strongly suggest temperature-independent inelastic scattering in this sample, leading to an additional reduction of the coherence length.

As it has been already mentioned, both weak localization and Raman spectroscopy give information about distances between defects and it is worth comparing the results obtained from both techniques. By fitting the magnetotransport data to a theoretical model of weak localization in graphene, we obtained characteristic scattering lengths. On the other hand, using the intensity ratio of the G peak to the D peak ($R_{GD}$), it is possible to estimate the average grain diameter ($L_\alpha$):[9,38]

$$L_\alpha(nm) = 2.4 \cdot 10^{-10} \lambda^4 R_{GD}, \qquad (8)$$

and the average distance between the defects ($L_D$) :[10]

$$L_D(nm) = \sqrt{1.8 \cdot 10^{-9} \lambda^4 R_{GD}}, \qquad (9)$$

where $\lambda$ is the wavelength of the exciting laser light in nanometers. Accounting for the experimentally obtained maxima of the $R_{GD}$ ratio distribution, grain diameters and distances between defects for both samples were calculated. These values were compared with the elastic scattering lengths and the decoherence length extrapolated to 300 K, obtained from the weak localization measurement. The results are collected in Table 3.

Table 3. Grain diameter ($L_\alpha$) and distance between defects ($L_D$) calculated using eq. (5) and (6) respectively from maxima of $R_{GD}$ ratio distributions (figure 5a and 5d). Intervalley scattering length ($L_i$) and coherence length extrapolated to 300 K ($L_\varphi$ @300 K) from weak localization fit for graphene deposited on epilayer, graphene deposited on NWs and their ratios.

| Graphene on: | GaN NWs | EPI GaN | | Ratio | |
|---|---|---|---|---|---|
| Maximum label | d1 | a1 | a2 | a1/d1 | a2/d1 |
| $L_\alpha$ (nm) | 7 | 21 | 59 | 3 | 8.4 |
| $L_D$ (nm) | 7 | 12 | 21 | 1.7 | 3 |
| $L_i$ (nm) | 125 | 884 | | 7.1 | |
| $L_\varphi$ @ 300 K (nm) | 32 | 53 | | 1.7 | |

Although the exact values of lengths obtained from both techniques are significantly different, the correlation between them can be showed when looking at their ratios. As expected, the ratio of the grain diameter obtained from the $R_{GD}$ ratio distribution maximum (d1) for graphene deposited on the NWs and the second maximum (a2) for graphene deposited on the epitaxial layer correlate very well with the ratio of the intervalley scattering lengths. Therefore, we can conclude that the reduction of the intervalley scattering length for graphene on the NWs can be understood as an influence of the creation of the hopping defects, as shown in the Raman measurements. The hopping defects can create a way for the electron to scatter between different sublattices in graphene and change its isospin. On the other hand, the ratio of the distance between defects obtained from the maximum (d1) for graphene on the NWs and the first maximum (a1) for graphene on the epilayer shows a correlation with the ratio of the coherence length extrapolated to 300 K. Therefore, a decrease of the coherence length for graphene on the NWs can be explained by scattering on randomly distributed spin carrying defects. This would suggest the involvement of coulomb defects, induced in the NWs/graphene interface by the spontaneous and piezoelectric polarization of GaN, which can give a substantial contribution to the Raman signal due to a very small distance from the graphene sheet (a few angstroms).

## IV. CONCLUSIONS

In this paper, a significant enhancement of electron scattering in graphene deposited on the GaN NWs was studied using Raman spectroscopy and microwave-induced contactless electron transport. It was found that Raman spectra are strongly correlated with nanowire distribution underneath the graphene layer. The modulation of the electron concentration, observed together with a homogenous strain induced in graphene deposited on the GaN NWs, suggests that the Raman enhancement could be explained in terms of a strong local electric field induced by electric charges located on the nanowire tips, thus similarly to the mechanism proposed in Surface Enhancement Raman Scattering. The observed modulation of the carrier concentration resembles the distribution of the surface charge density present on the NW tips due to the spontaneous and piezoelectric effect. A detailed analysis of the enhancement ratio for individual Raman bands shows the contribution of the Tip Enhancement Raman Scattering (TERS) mechanism and additional scattering on defects has to be taken into account. For graphene deposited on the GaN epitaxial layer, a substantial, yet lower Raman enhancement was observed almost exclusively on the microstep edges. This effect can be explained in terms of the influence of localised polarization charges existing near the GaN microstep edges, which is not as strong as the one observed in graphene deposited on the NWs. Most probably, electric charges concentrated on top of the GaN NWs give a higher contribution to the Raman enhancement than charges located both on the terraces and the macrostep edges. A careful analysis of the intensity ratio of the G and D peaks and the intensity ratio of the D and D' peaks showed that the nanowires introduce a homogenous defect density in graphene in contrast to the GaN epitaxial layer, where defects in graphene are distributed mainly near the microstep edges of GaN. A statistical analysis of these ratios showed substantial differences in both distances between defects and defect types for both samples. The study of the intensity ratio of the D and D' peaks suggested three types of defects. A mixture of vacancies, vacancies and hopping defects were observed for graphene deposited on the NWs. In contrast, we found six types of defects, including on-site defects, graphene boundaries, a mixture of vacancies, vacancy defects, hopping defects as well as $sp^3$ hybridization defects for graphene deposited on the GaN epilayer. Interestingly, the probability of each defect type maximum is definitely different for graphene on the epilayer and graphene on the NWs. The observed difference in the defect density for both samples can be understood as a combination of the activation and deactivation of some types of defects induced by the surface influence.

Weak localization measurements showed the reduction of the elastic intervalley scattering length for graphene deposited on the NWs as compared to graphene deposited on the epilayer, which we correlated with the creation of hopping defects observed in the Raman measurements. The temperature dependence of the coherence length showed electron-electron scattering in a diffusive regime as the main decoherence mechanism for both samples, with the additional temperature-independent scattering for graphene deposited on the NWs, which we explained by scattering on charges induced on the NWs/graphene interface by the spontaneous and piezoelectric polarization of GaN. Both experimental techniques gave consistent results as regards the scattering process and the nature of defects in graphene deposited on the GaN NWs.

Our results demonstrate an excellent research potential of graphene deposited on the GaN NWs. The possibility of modulating the carrier concentration with a constant strain is crucial during the fabrication of transistors and sensors. The homogenous defect distribution is important in both solar cells and sensor applications. More investigations are needed to explain the mechanism of the Raman enhancement. In the future the influence of different nanowire diameters and materials on the observed phenomena is well worth studying.

## ACKNOWLEDGMENTS

This work was partially supported by the NCN grant no 2012/07/B/ST3/03220, Poland, NCBiR grant GRAF TECH/NCBR/02/19/2012, Poland and by Polish Ministry of Science and Higher Education in years 2015-2019 as a research grant "Diamentowy Grant".. ZRZ is grateful for support from the grant Innovative Economy POI G.01.03.01 00-159/08 (InTechFun).

___

*To whom correspondence should be addressed. E-mail: jakub.kierdaszuk@student.uw.edu.pl